\documentclass[aps,preprint,superscriptaddress,showpacs,amsmath,amssymb,endfloats*]{revtex4}

\usepackage{graphicx}
\usepackage{bm}

\begin{document}

\title{Theoretical analysis of a single and double reflection atom
interferometer in a weakly-confining magnetic trap}
\author{James A. Stickney}
\affiliation{Department of Physics, Worcester Polytechnic Institute,
100 Institute Road, Worcester, Massachusetts 01609, USA}
\author{Rudra P. Kafle}
\affiliation{Department of Physics, Worcester Polytechnic Institute,
100 Institute Road, Worcester, Massachusetts 01609, USA}
\author{Dana Z. Anderson}
\affiliation{Department of Physics and JILA, University of Colorado
and National Institute of Standards and Technology, Boulder,
Colorado 80309-0440, USA}
\author{Alex A. Zozulya}
\affiliation{Department of Physics, Worcester Polytechnic Institute,
100 Institute Road, Worcester, Massachusetts 01609, USA} \email[]{
zozulya@wpi.edu}

\begin{abstract}
The operation of a BEC based atom interferometer, where the atoms
are held in a weakly-confining magnetic trap and manipulated with
counter-propagating laser beams, is analyzed.  A simple analytic
model is developed to describe the dynamics of the interferometer.
It is used to find the regions of parameter space with high and low
contrast of the interference fringes for both single and double
reflection interferometers. We demonstrate that for a double
reflection interferometer the coherence time can be increased by
shifting the recombination time. The theory is compared with recent
experimental realizations of these interferometers.
\end{abstract}


\maketitle

\section{Introduction}

A promising method for building an atom interferometer has been
demonstrated by several groups \cite{wang05,garcia06, wu05}. This
method uses a standing light wave to manipulate a Bose-Einstein
condensate (BEC) that is confined in a waveguide with a weak
trapping potential along the guide.

\begin{figure}
\includegraphics[width=8.6cm]{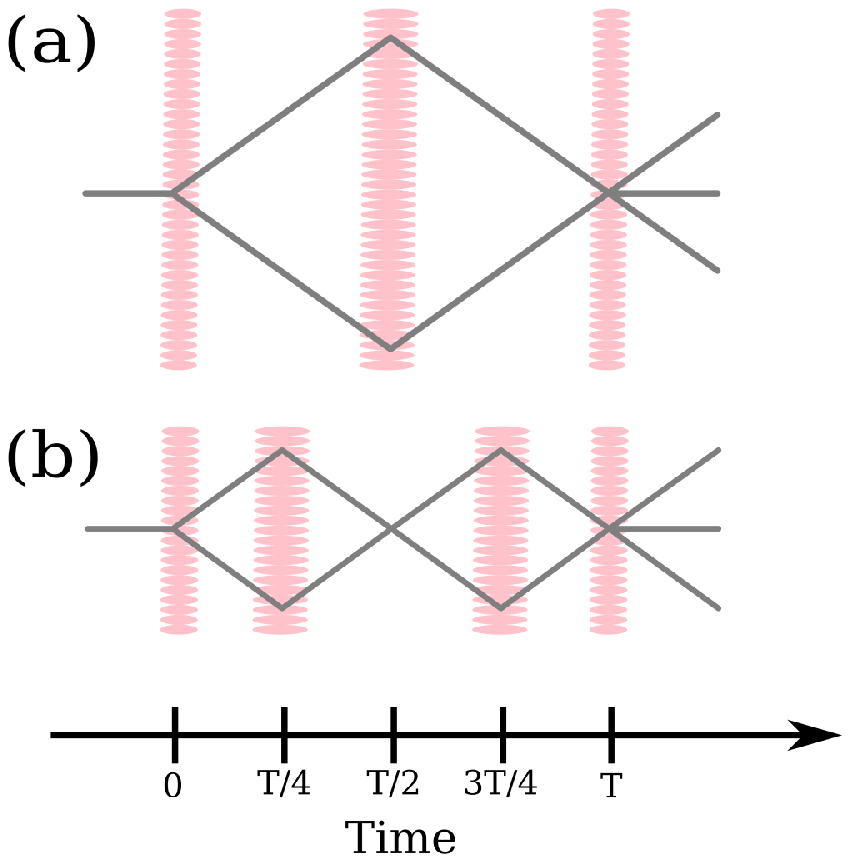}%
\caption{(color online) Trajectories of the BEC clouds as functions
of time for (a) a single reflection interferometer and (b) a double
reflection interferometer. Vertical wavy bands show timing of the
optical pulses. \label{cartoon}}
\end{figure}

The trajectories of the BEC clouds during the interferometric cycle
are shown in Fig.~\ref{cartoon} (a). The cycle of duration $T$
starts at $t=0$ by illuminating the motionless BEC with the wave
function $\psi_{0}$ with a splitting pulse from the two
counterpropagating laser beams. This pulse acts like a diffraction
grating splitting the initial BEC cloud into two harmonics
$\psi_{+}$ and $\psi_{-}$. The atoms diffracted into the $+ 1$ order
absorb a photon from a laser beam with the momentum $\hbar k_{l}$
and re-emit it into the beam with the momentum $-\hbar k_{l}$
acquiring the net momentum of $2\hbar k_{l}$. The harmonic
$\psi_{+}$ starts moving with the velocity $v_{0} = 2\hbar k_{l}/M$,
where $k_{l}$ is the wavenumber of the laser beams and $M$ is the
atomic mass. Similarly, the harmonic $\psi_{-}$ starts moving with
the velocity $-v_{0}$. The two harmonics are allowed to propagate
until the time $t=T/2$. At this time the harmonics are illuminated
by a reflection optical pulse. The atoms in the harmonic $\psi_{+}$
change their velocity by $-2v_{0}$ and those in the harmonic
$\psi_{-}$ by $2v_{0}$. The harmonics propagate until the time $t=T$
and are subject to the action of the recombination optical pulse.
After the recombination, the atoms in general populate all three
harmonics $\psi_{0}$ and $\psi_{\pm}$. The relative population of
the harmonics depend on the phase difference between the harmonics
$\psi_{\pm}$ acquired during the interferometric cycle.  By counting
the number of atoms in each harmonic the phase difference can be
determined.  This type of an interferometer will be referred to as a
single reflection interferometer.

The first experiments using this type of interferometer were done by
Wang et al. \cite{wang05}. Good contrast of the interference fringes
was observed for cycle times not exceeding about 10~ms. This fact
was theoretically explained by Olshanii et al. \cite{olshanii05}.
The authors of Ref.~ \cite{olshanii05} attributed the loss of
contrast to a distortion of the phase across each harmonic that was
caused by both the atom-atom interactions and the residual potential
along the waveguide.

\begin{figure}
\includegraphics[width=8.6cm]{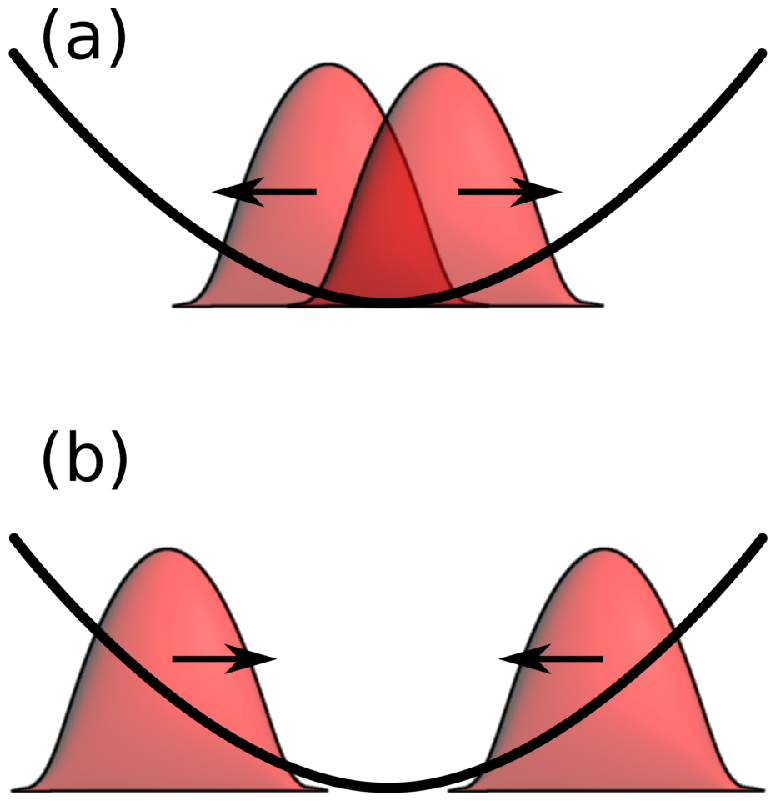}%
\caption{(color online) The two different forces acting on the two harmonics
during the interferometric cycle. (a) When the two harmonics overlap,
the atom-atom interactions cause a repulsive force between the
harmonics and (b) the external potential exerts a force pushing the two harmonics
towards the center of the trap. \label{Forces}}
\end{figure}

A simple way to understand the reason for the loss of contrast is to
consider the effect of the two forces acting on the harmonics
$\psi_{\pm}$ during the interferometric cycle. The first force is
due to a repulsive nonlinearity between the two BEC clouds, as shown
in Fig.~\ref{Forces}~(a). This force exists only when the two
harmonics overlap. The second force is exerted by the trapping
potential along the waveguide and pushes the harmonics toward the
center of the trap as shown in Fig.~\ref{Forces} (b). As a result,
the velocities of the harmonics are not equal to their initial
values $\pm v_0$ during the interferometric cycle.

Consider, for example, the $\psi_{+}$ harmonic. Just before the
reflection pulse its velocity is $v_0 + \delta v$. Since the
reflection pulse changes the speed of each harmonic by $2 v_0$, just
after the reflection pulse the velocity of the $\psi_{+}$ harmonic
becomes $-v_0 + \delta v$. Because of this fact, when the harmonic
moves back to the center of the trap, its velocity before the
recombination is not equal to $-v_0$. The change in the speeds of
the harmonics during the first and the second halves of the
interferometric cycle has two consequences. First, harmonics do not
completely overlap at the nominal recombination time. Second, the
recombination optical pulse is unable to exactly cancel the
harmonics' velocities causing the recombined wave functions to have
coordinate-dependent phases across the clouds. Both mechanisms
result in a washout of interference fringes and loss of contrast.

A modification of the single-pass interferometer interferometer
shown in Fig.~\ref{cartoon} (b), was built by Garcia et al.
\cite{garcia06}. The interferometric cycle begins by illuminating
the BEC with a splitting pulse. The two harmonics freely propagate
until the time $t=T/4$ when they are illuminated by a reflection
pulse. They continue to freely propagate until the time $t = 3T/4$
when a second reflection pulse is applied. The harmonics freely
propagate until the time $t = T$ when they overlap and are subject
to the action of the recombination pulse. This type of an
interferometer will be referred to as a double reflection
interferometer. The coherence time demonstrated in Ref. \cite{garcia06}
exceeded 44~ms.  More recently, the coherence time of this interferometer has been extended to over 80~ms \cite{burke07}.
The advantage of a double reflection interferometer over a the
single reflection one is in that the shift in the velocity of the
harmonics is considerably reduced allowing for larger cycle times.
Recently this type of interferometer was used to measure the ac
Stark shift \cite{deissler07}.

Along with the experimental realization of the single and double
reflection interferometer, the authors of Ref. \cite{burke07}
developed a theoretical model to describe its operation. The model
presented in this paper differs from that of Ref. \cite{burke07} by
accounting for the effects of atom-atom interactions, the change in
the BEC size during the cycle time, and the incomplete overlap of
the two harmonics at the recombination time. These effects do not
significantly change the results of analysis of Ref. \cite{burke07}
for the single reflection interferometer. However, our results for
the double reflection interferometer are of a different functional
form than those found in Ref. \cite{burke07}.

In the rest of the paper, we use a simple analytic model to
calculate the momentum and the degree of overlap of the two
harmonics at the end of the cycle for both the single and double
reflection interferometers. Both of these depend on the time that
the two harmonics spend overlapping, the total cycle time and the
frequency of the trap. Next, we find the regions of a large and a
small interference fringe contrast for both the single and double
reflection interferometers. We demonstrate that, with a double
reflection interferometer, the coherence time can be increased by
shifting the recombination time. Finally, we compare the model with
recent experimental realizations of these interferometers.

\section{Analytical model} \label{Model}
The dynamics of a BEC in a waveguide will be analyzed in the
framework of the Gross-Pitaevskii equation (GPE)
\begin{equation}\label{GPE}
i \frac{\partial}{\partial \tau} \psi(x, \tau) = \left[ -
\frac{1}{2} \frac{\partial^2}{\partial x^2} + v(x) + \Omega(\tau)
\cos x + p|\psi|^2
 \right] \psi(x,\tau),
 \end{equation}
which has been obtained by projecting the three-dimensional GPE onto
the strongly confining transverse mode of the waveguide (for more
details see \cite{stickney07}). In Eq.~(\ref{GPE}), $\psi$ is the
wave function of the BEC normalized to one, the dimensionless
coordinate $x$ is measured in units of $x_0 = 1 / 2 k_l$ and the
dimensionless time $\tau$ is measured in units of $t_0 = M / 4 \hbar
k_l^2$, where $k_l$ is the wave vector of the lasers  and $M$ is the
atomic mass.

The weakly confining potential along the guide is harmonic
\begin{equation}\label{potential}
    v(x) =  \frac{1}{2}\omega^{2} x^{2},
\end{equation}
and $\Omega(\tau) \cos x$ is the potential associated with the laser
beams. The strength of interatomic interaction is given by the
parameter
\begin{equation}
p = a_s N / a_\perp^2 k_l, \label{Nonlinearity}
\end{equation}
 where $a_s$ is the s-wave
scattering length, $N$ is the total number of atoms in the BEC,
$a_\perp = \sqrt{\hbar / M \omega_\perp}$ is the transverse harmonic
oscillator length, and $\omega_{\perp}$ is the transverse frequency
of the guide.

The optical potential $\Omega(\tau) \cos x$ acts as a diffraction
grating for the BEC wave function $\psi$. This grating diffracts the
BEC into several harmonics separated by multiples of the grating
wave vector:
\begin{equation} \label{WaveFunction}
\psi = \sum_n \psi_n e^{i n x},
\end{equation}
where $\psi_n$ are the slowly-varying amplitudes of the harmonics'
wave functions.

The dynamics of the BEC due to the optical potential has been
discussed in Refs.~\cite{wu05, garcia06, stickney07}. Since the
optical pulses used to manipulate the BEC are intense and short
their action can be described by simple mixing matrices operating on
the harmonics $\psi_{n}$ in Eq.~(\ref{WaveFunction}). The splitting
pulse transforms the initial zero-momentum harmonic $\psi_{0}$ into
the two harmonics with $n = \pm 1$: $\psi_0 \rightarrow (\psi_{+1} +
\psi_{-1})/\sqrt{2}$. The reflection pulse transforms the $n = \pm
1$ harmonic into the $n = \mp 1$ harmonic: $\psi_{\pm 1} \rightarrow
\psi_{\mp 1}$. Finally, after the recombination pulse, the BEC
consists of three harmonics with $n = -1, 0, +1$. The population in
the zero momentum harmonic $\psi_0$ depends on the relative phase
shift of the $\pm 1$ harmonics immediately before the recombination.

Between the splitting and the recombination optical pulses, the BEC
consists of two harmonics with $n = \pm 1$. In the Thomas-Fermi
approximation, the evolution of these harmonics is governed by the
set of equations
\begin{eqnarray}\label{TF_limit}
    &&\left(\frac{\partial}{\partial \tau} \pm \frac{\partial}{\partial
    x}\right)n_{\pm} = - \frac{\partial}{\partial x}\left(
    n_{\pm}\frac{\partial \phi_{\pm}}{\partial x}\right),\nonumber \\
    &&\left(\frac{\partial}{\partial \tau} \pm \frac{\partial}{\partial
    x}\right)\phi_{\pm} = - \frac{1}{2}\left(\frac{\partial \phi_{\pm}}{\partial
x}\right)^{2} -
    v - p(n_{\pm} + 2n_{\mp}),
\end{eqnarray}
where $n_{\pm}$ and $\phi_{\pm}$ are densities and phases of the
harmonics introduced by the relations
\begin{equation}\label{density_and_phase}
\psi_{\pm 1} = \sqrt{n_{\pm}}\exp(i\phi_{\pm}).
\end{equation}
Equations (\ref{TF_limit}) are valid when $p R \gg 1$, where $p$ is
the dimensionless nonlinearity parameter and $R$ is the
characteristic size of the harmonics.

The set of partial differential equations Eqs.~(\ref{TF_limit}) can
be transformed into a set of ordinary differential equations by
parametrizing the density and phase of the harmonics as
\begin{eqnarray}  \label{TF_density_and_phase}
    n_{\pm} &=& \frac{3}{8R}\left[1 -
    \frac{(x-x_{\pm})^{2}}{R^{2}}\right], \nonumber \\
    \phi_{\pm} &=& \varphi_{\pm} + \kappa_{\pm}(x-x_{\pm}) +
    \frac{g}{2}(x-x_{\pm})^{2} + \frac{1}{6} s_{\pm} (x - x_{\pm})^3.
\end{eqnarray}
Functions $R$, $x_\pm$, $\varphi_\pm$, $\kappa_\pm$, $g$, and
$s_\pm$ depend only on time.  The radius of each of the harmonics is
$R$. The position of each harmonics's center of mass is given by the
coordinate $x_\pm$. The coordinate-independent part of the phase of
each harmonic is $\varphi_\pm$. The correction to the wave vector of
each harmonic is $\kappa_\pm$ and the curvature of the phase is
given by the parameter $g$. Finally, the parameter $s_\pm$
determines the size of the cubic contribution to the phase.

Using Eqs.~(\ref{TF_density_and_phase}) in Eqs.~(\ref{TF_limit})
results in a set of ordinary differential equations for the
parameters $R$, $x_\pm$, $\varphi_\pm$, $\kappa_\pm$, $g$. These
equations are
\begin{eqnarray}\label{final_set_of_parab_eqns}
    &&R^{\prime} = g R, \nonumber \\
    && x_{\pm}^{\prime} = \pm 1 + \kappa_{\pm}, \nonumber \\
    && \kappa_{\pm}^{\prime} =  - \omega^{2}  x_{\pm} \pm \frac{\omega^2 R_{0}^3}{4 R^{2}}d_{1}(q), \nonumber \\
    &&g^{\prime} = - g^{2} - \omega^{2} +
    \frac{\omega^2 R_{0}^3 }{2 R^{3}}[1 + d_{2}(q)], \nonumber \\
    && s'_{\pm} = \frac{6}{2} \frac{\omega^2 R_{0}^3}{2R^4}
    d_{3}(q), \nonumber \\
    && \varphi_{\pm}^{\prime} = \frac{1}{2} \kappa_{\pm}^{2} -
    \frac{1}{2} \omega^{2} x_{\pm}^{2} + \frac{\omega^2 R_{0}^3}{4 R} [1 +
    d_0(q)].
\end{eqnarray}
Here
\begin{eqnarray}\label{ds}
    && d_{0} =  \left( 2 - \frac{7}{2} |q|^{2} + 2 |q|^{3} - \frac{1}{8} |q|^{5}  \right)\theta(|q| < 2),
\nonumber \\
    &&d_{1} = q\left(4 + \frac{15}{2} |q| - \frac{35}{2} |q|^2 + \frac{65}{8}
|q|^3 - \frac{7}{16} |q|^5  \right)\theta(|q| < 2),
    \nonumber \\
    &&d_{2} = \left(2 - \frac{15}{2}|q|^{2} + 5|q|^{3} -
\frac{3}{8}|q|^{5}\right)\theta(|q| < 2), \nonumber \\
    &&d_{3} = q \left(
-\frac{35}{2} |q| + \frac{175}{6} |q|^2 - \frac{105}{8} |q|^3 + \frac{35}{48}
|q|^5
\right)\theta(|q| < 2),
\end{eqnarray}
and  $q = (x_{+} - x_{-}) / R$ is the relative displacement of the
two harmonics.   The $\theta$ function in Eq.~(\ref{ds}) is equal to
one if its argument is a logical true and zero if it is a logical
false. The nonlinearity parameter $p$ was eliminated from
Eqs.~(\ref{final_set_of_parab_eqns}) with the help of the relation
\begin{equation}
 p = \frac{2}{3}\omega^{2}R_{0}^{3}, \label{R0}
\end{equation}
where $R_0$ is the initial size of the BEC cloud (equal to the size
of the both harmonics immediately after the splitting pulse).
Equation (\ref{R0}) assumes that the BEC is created in the confining
potential Eq.~(\ref{potential}).

The procedure of deriving Eqs.~(\ref{final_set_of_parab_eqns})
parallels that given in \cite{stickney07}.
Equations~(\ref{final_set_of_parab_eqns}) and (\ref{ds}) differ from
those found in Ref. \cite{stickney07} by accounting for an
additional cubic term.

Since the BEC is in the lowest stationary state of the trap before
the splitting pulse, the initial conditions for
Eqs.~(\ref{final_set_of_parab_eqns}) are $R(\tau=0) = R_{0}$, and
$x_{\pm}(\tau=0) = \kappa_\pm(\tau=0) = g(\tau=0) = \varphi_\pm(\tau
= 0) = 0$.  The reflection pulses are accounted for by the boundary
conditions at the time of the reflections: $x_\pm \rightarrow x_\mp$
and $\kappa_\pm \rightarrow \kappa_\mp$.

After the recombination pulse the BEC consists of three harmonics
$\psi_0$ and $\psi_{\pm}$. The population of the zero-momentum
harmonic is given by the expression
\begin{equation}
{N_{0}} = \frac{1}{2} \left[1 + V \cos \Delta \varphi \right],
\end{equation}
where $\Delta \varphi = \varphi_{+} - \varphi_{-}$ is the relative
phase difference between the harmonics.

The fringe contrast $V$ is given by the relation
\begin{equation}\label{contrast}
    V =  \frac{3}{2}\int_{0}^{1 - |\Delta x|/2 R} dy
    \left[ \left(1 - y^2 + \frac{(\Delta x)^2}{4 R^2} \right)^2  -
\frac{(\Delta x)^2}{R^2}\right]^{1/2}\cos ( \Delta k R y +
\frac{1}{6} \Delta s R^3 y^3 ).
\end{equation}
Here
\begin{equation}\label{DeltaK}
\Delta k = \Delta \kappa - g \Delta x +  \frac{1}{8} \Delta s
(\Delta x)^2,
\end{equation}
$\Delta x = x_+ - x_-$, $\Delta \kappa = \kappa_+ - \kappa_-$ and
$\Delta s = s_+ - s_-$.

When the fringe contrast is high ($1 - V \ll 1$ ),
Eq.~(\ref{contrast}) can be simplified to
\begin{eqnarray} \label{contrast_final}
 V &\approx& 1 -
\frac{3}{2} \left(\frac{\Delta x}{2 R} \right)^2 \left[ \ln\left|\frac{2 R}{\Delta x}\right| + 2 \ln 2 - \frac{1}{2}\right] \nonumber \\
&-& \frac{1}{10} \left(\Delta k R + \frac{1}{14} \Delta s R^3
\right)^2 - \frac{1}{6615}( \Delta s R^3)^2 \nonumber \\
&=& 1 - A - B - C,
\end{eqnarray}
where $0 \le A,B,C \ll 1$.

The expression for the fringe contrast $V$ given by
Eq.~(\ref{contrast_final}) contains three terms $A$, $B$, and $C$.
All these terms are positive and decrease the fringe contrast
additively. In the following analysis their influence will be
considered separately. The boundary between the regions of high and
low fringe contrast will be defined by the conditions $A \sim 1/2$,
or $B
 \sim 1/2$, or $C \sim 1/2$. Despite the fact that
 Eq.~(\ref{contrast_final}) was obtained in the limit $A,B,C \ll 1$,
 these conditions turn out to be good qualitative and quantitative
 approximations.

The term
\begin{equation} \label{A}
 A = \frac{3}{2} \left(\frac{\Delta x}{2 R} \right)^2 \left[ \ln\left|\frac{2 R}{\Delta x}\right| + 2 \ln 2 - \frac{1}{2}\right]
\end{equation}
describes the decrease in the fringe contrast due to the incomplete
overlap of the two harmonics at the recombination time. The region
of low contrast due to the incomplete overlap is given by the
condition
\begin{equation}
|\Delta x|/ R \gtrsim 1 \label{xpoR}
\end{equation}
(the harmonics overlap by less than half of their their full
widths).

The term
\begin{equation} \label{B}
 B = \frac{1}{10} \left(\Delta k R + \frac{1}{14} \Delta s R^3 \right)^2
\end{equation}
describes the loss of fringe contrast due to the phase difference
between the center and the periphery of the cloud $\psi_0$. The
region of low fringe contrast is given by the relation
\begin{equation}
\left|\Delta k R + \frac{1}{14} \Delta s R^3\right| \gtrsim \sqrt{5}
\sim 2. \label{DkR}
\end{equation}

The phase difference in Eq.~(\ref{B}) is due to a combination of
both the quadratic and the cubic terms in the expression for the
phase (\ref{TF_density_and_phase}). These terms have been grouped
together in Eq.~(\ref{B}) because it is sometimes possible to set
$B=0$ by shifting the recombination time, as will be shown in
Sec.~\ref{Double_Reflect}. However, even when $B=0$, there are
higher order phase distortions that can cause a loss of fringe
contrast due to the presence of the cubic term in the phase
Eq.~(\ref{TF_density_and_phase}). The effect of this cubic term on
the fringe contrast is given by the parameter
\begin{equation} \label{C}
 C = \frac{1}{6615}( \Delta s R^3)^2.
\end{equation}
This term results in a small fringe contrast when
\begin{equation}
\left|\Delta s\right| R^3 \gtrsim \sqrt{6615/2} \sim 60. \label{sR3}
\end{equation}

Each of the three above-discussed contributions to the
Eq.~(\ref{contrast_final}) can be expressed in terms of three
dimensionless parameters having a simple physical meaning. The first
parameter is the dimensionless trapping frequency
\begin{equation}
    \omega = \omega_{\parallel}t_{0}, \label{omega}
\end{equation}
The second parameter is the product of the interferometric cycle
time and the trapping frequency
\begin{equation}
    \omega T  = \omega_{\parallel} T_D \label{omegaT}
\end{equation}
where $\omega_{\\}$ and $T_{D}$ are the dimensional trap frequency
and the cycle time, respectively.

The third parameter is the product of the dimensionless initial size
of the BEC $R$ and the trapping frequency $\omega$. This parameter
can be written in terms of dimensional quantities as
\begin{equation}
 \omega R = \frac{\omega_{\parallel} R_{D} }{v_0} , \label{omegaR}
\end{equation}
where $R_D$ is the dimensional radius of the BEC and $v_0 = 2 \hbar
k_{l} / M$ is the speed of the harmonics just after the splitting
pulses.  This parameter is the time it takes the two harmonics to
separate measured in units of the inverse trapping frequency.

\section{Single reflection interferometer} \label{Single_Reflect}
Solutions of Eqs.~(\ref{final_set_of_parab_eqns}) for the case of a
single reflection interferometer shown in Fig.~\ref{cartoon} (a)
have been previously discussed in \cite{stickney07} without the
cubic phase term. Inclusion of this term ($\Delta s$) is somewhat
cumbersome but straightforward and results in the following
expressions for $R$, $\Delta x$, $\Delta \kappa$, $g$ and $\Delta s$
at the end of the interferometric cycle of duration $T$:
\begin{eqnarray}
R &=& R_0 [ 1 - \frac{1}{4} (\omega T)^2], \nonumber \\
\Delta x &=&  \frac{1}{4\omega} (\omega T)^3, \nonumber \\
\Delta \kappa &=& \frac{1}{2} (\omega T)^2 -  2 (\omega R_0)^2 D_1(\omega T/\omega
R_0), \nonumber \\
g &=& - \omega [\omega T - 2  \omega R_0 D_2(\omega T/\omega R_0) ], \nonumber \\
\Delta s &=& 2 \omega^2 D_3( \omega T/\omega R_0).
\label{SingleReflect}
\end{eqnarray}
Here
\begin{equation} \label{D1}
D_1(x) = \left\{
\begin{array}{ccc}
\frac{1}{4} x^2 (
2 + \frac{5}{2} x - \frac{35}{8}x^2 + \frac{13}{8} x^3 - \frac{1}{16}x^5
) &,& x<2 \\
1/2 &,& x > 2
\end{array}
\right. ,
\end{equation}
\begin{equation}\label{D2}
    D_{2}(x) = \left\{\begin{array}{ccc}
    \frac{1}{4} x\left(2 - \frac{5}{2}x^{2} + \frac{5}{4}x^{3} -
    \frac{1}{16}x^{5}\right) &,& x < 2, \\
     0 &,& x > 2.
    \end{array}\right. ,
\end{equation}
and
\begin{equation}\label{D3}
    D_{3}(x) = \left\{\begin{array}{ccc}
    \frac{3}{2} x^3\left(
\frac{35}{6} - \frac{175}{24} x + \frac{21}{8}x^2 - \frac{5}{48} x^4
\right) &,& x < 2, \\
     1 &,& x > 2.
    \end{array}\right.
\end{equation}
In deriving the above expressions, only the lowest order
contributions in terms of $\omega T$ and $\omega R$ were retained.
The loss of the fringe contrast takes place when both these
parameters are still small. The first of Eq.~(\ref{SingleReflect})
then shows that the relative change in the size of the BEC during
the cycle is small and will be neglected in the subsequent analysis.

The loss of the contrast due to the term $A$ Eq.~(\ref{A}) happens
for $A
> 1/2$. Using Eqs.~(\ref{xpoR}) and (\ref{SingleReflect}), we can translate this inequality
into the relation
\begin{equation}
 \frac{1}{4} \frac{(\omega T)^3} {\omega R} > 1. \label{Dx_single}
\end{equation}
Similarly, the loss of the contrast due to the term $B$
Eq.~(\ref{B}) with the help of Eqs.~(\ref{DkR}) and
(\ref{SingleReflect}) can be expressed as
\begin{equation}
\left\{\frac{1}{2} (\omega T)^2 - 2 (\omega R)^2  \left[ D_1(\omega
T/ \omega R) - \frac{1}{14} D_3(\omega T / \omega R) \right]
 \right\} \frac{\omega R}{\omega}  > 5, \label{DkR_single}
\end{equation}
where $D_1$ and $D_3$ are given by Eqs. (\ref{D1}) and (\ref{D3})
respectively.

\begin{figure}
\includegraphics[width=8.6cm]{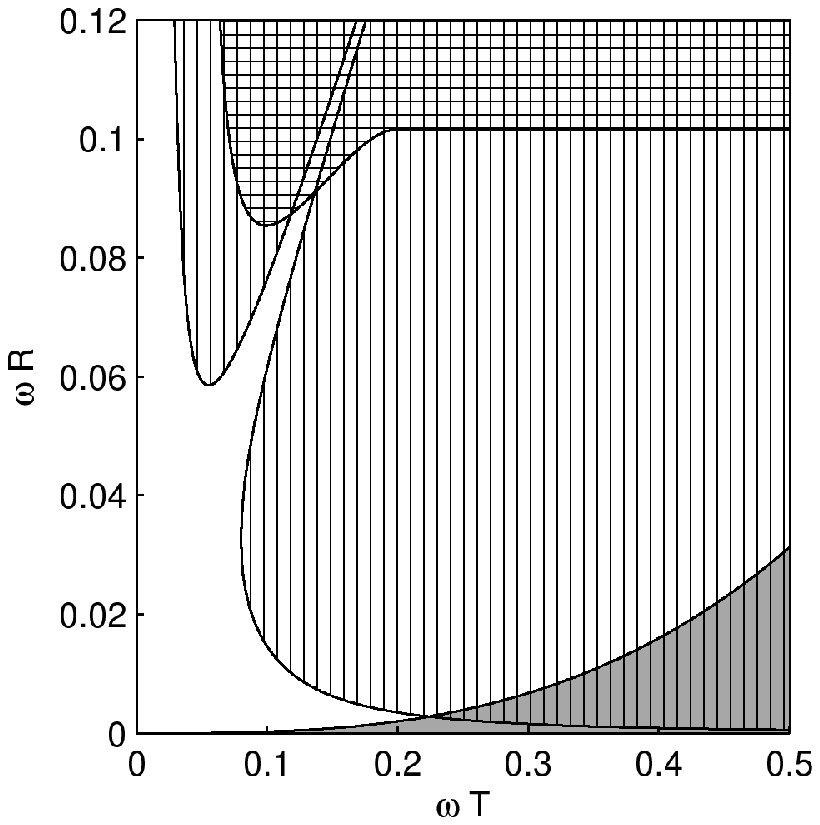}%
\caption{\label{Nominal_Map_Single1} The regions of large and small
fringe contrast for the single reflection interferometer. The white
region corresponds to high contrast. In the grey region $A
> 1/2$, the region filled with vertical stripes is where $B > 1/2$.
The region filled with horizontal stripes is where $C > 1/2$. }
\end{figure}

Finally, the loss of the fringe contrast due to the term $C$
corresponds to the region of parameters where
\begin{equation}
 \frac{(\omega R)^3}{\omega} D_3(\omega T / \omega R) > 30. \label{DsR3_single}
\end{equation}
 Figure~\ref{Nominal_Map_Single1} is a two-dimensional plot showing
the regions of operation of the interferometer. The dimensionless
trap frequency is $\omega = 3.5 \times 10^{-5}$, which roughly
corresponds to the value used in recent experiments \cite{burke07}.
The white region corresponds to large fringe contrast. In the grey
region (lower right corner) $A > 1/2$ and the fringe contrast is
small because of the lack of overlap of the harmonics $\psi_\pm$.
The region with the vertical stripes corresponds to $B > 1/2$ and
the contrast is small because of the phase difference across the
cloud. The region filled with the horizontal stripes corresponds to
$C > 1/2$ and the contrast is lost because of large value of the
cubic phase across the harmonic. These regions were found by
numerically inverting Eqs. (\ref{Dx_single}), (\ref{DkR_single}),
and (\ref{DsR3_single}). Note that when two shaded regions overlap,
the contrast is lost due to two different mechanisms.

Figure \ref{Nominal_Map_Single1} shows that when $\omega R < 5
\times 10^{-3}$, the contrast is lost because the two harmonics
$\psi_\pm$ do not overlap at the recombination time. In the region
$5 \times 10^{-3} < \omega R < 6 \times 10^{-2}$, the contrast is
lost because the trap causes a difference in the phase across the
atomic cloud. When $\omega R
> 6 \times 10^{-3}$ and $\omega T < \sqrt{2} \omega R$, the force that the two
harmonics exert on each other is the cause of the difference in
phase across the harmonics. For the cycle time $\omega T = \sqrt{2}
\omega R$, there is no phase difference across the harmonics, but
for $\omega T > 0.12$ the cubic phase across the recombined harmonic
causes a loss of fringe contrast.  For $\omega R
> 0.06$ and $\omega T > \sqrt{2} \omega R$ the trap causes the loss
of fringe contrast.

The boundary between the regions where $A<1/2$ and $A>1/2$ is given
by the relation
 \begin{equation}
 \omega R = \frac{1}{4} (\omega T)^3.\label{Lower_Single}
\end{equation}
For a given value of $\omega T$, Eq.~(\ref{Lower_Single}) sets the
lower limit on the parameter $\omega R$ for which the fringe
contrast is large.

The boundary between the regions where $B < 1/2$ and $B > 1/2$
obtained with the help of Eq.~(\ref{DkR}) in the limit $\omega R \ll
1$, is given by the relation
\begin{equation}
 \omega R = \frac{4 \omega }{(\omega T)^2}. \label{Upper_Single}
\end{equation}
For a given $\omega T$, Eq.~(\ref{Upper_Single}) is the upper limit
on the parameter $\omega R$ for which the fringe contrast is large.

The largest cycle times with high fringe contrast correspond to the
point on Fig.~\ref{Nominal_Map_Single1} where the grey and vertical
striped regions meet. Eqs.~(\ref{Lower_Single}) and
(\ref{Upper_Single}) show that the maximum cycle time is given by
\begin{equation}
 \omega T_{max} = (16 \omega)^{1/5}
\end{equation}
and this occurs when the size of the BEC is given by the relation
\begin{equation}
 \omega R_{opt} = \frac{1}{4} ( 16 \omega )^{3/5}.
\end{equation}
The optimal value of the parameter $\omega R_{opt}$ is always
smaller than $\omega T_{max}$, which justifies the approximation
used in deriving Eq.~(\ref{Upper_Single}).

In a previous paper \cite{stickney07}, we demonstrated that it is
sometimes possible to increase the contrast of the interference
fringes by shifting the recombination time.  However, in the
geometry of a single-reflection interferometer when the BEC is
created in the confining potential given by Eq.~(\ref{potential}),
it is not possible to significantly increase the fringe contrast by
shifting the recombination time for $\omega T / \omega R > 2$, i.e.,
for the cycle times such that the two clouds completely separate. As
a result, the regions depicted in Fig. \ref{Nominal_Map_Single1}
cannot be significantly changed by shifting the recombination time.

\section{Double reflection interferometer} \label{Double_Reflect}

For the geometry of the double reflection interferometer shown in
Fig.\ref{cartoon}~(b), expressions for $R$ and $g$ were found  by
perturbatively solving Eqs.~(\ref{final_set_of_parab_eqns}) to third
order in $\omega T$ and first order in $\omega R$ yielding
\begin{eqnarray}
g &=& \omega\left[-\frac{1}{2}(\omega T) +\frac{1}{24}(\omega T)^3 + \frac{1}{4}(\omega R)D_2\left(\frac{\Delta x}{R}\right)\right],\nonumber \\
R &=& R_0 \left[1-\frac{1}{4}(\omega T)^2 \right].\label{R_g_final}
\end{eqnarray}
The equation for $R$ shows that, as for the single reflection
interferometer, the relative change in the size of the BEC during
the cycle is small. This change will be neglected in the subsequent
analysis.

Solutions of Eqs.~(\ref{final_set_of_parab_eqns}) results in the expressions for $\Delta x$ and $\Delta \kappa$ at the
end of the interferometric cycle that are given by
\begin{eqnarray}
\Delta x &=& \frac{2}{ \omega} \left[
2 \sin \frac{\omega T}{4} - 2 \sin \frac{3 \omega T }{4} + (1 + I_s ) \sin
\omega T
 \right], \nonumber \\
\Delta \kappa &=& 4 \cos \frac{\omega T}{4} - 4 \cos \frac{3 \omega
T}{4} + 2 (1 + I_s) \cos \omega T - 2 + 2 I_r. \label{x_k_d1}
\end{eqnarray}
Here
\begin{equation}
I_s = \frac{1}{4} (\omega R)^2 \label{Is}
\end{equation}
is the change in $\Delta \kappa$ caused by the repulsive force that
the two harmonics exert on each other during the separation and
\begin{equation}
I_r =\frac{1}{2} (\omega R)^2 \left[ D_1(|\Delta x|/R) - \frac{1}{2}
\right] \approx - \frac{1}{4} (\omega R)^2 \left[1 - \left( \frac{\Delta x}{R} \right)^2\right] \label{Ir}
\end{equation}
is the change in $\Delta \kappa$ caused by the force that the two
harmonics exert on each other during the recombination, with $D_1$
given by Eq.~(\ref{D1}).
Expanding Eq.~(\ref{x_k_d1}) into into a Taylor series and keeping up to the sixth order in $\omega T$
 results in the relations
\begin{eqnarray}
\Delta x &=& \frac{2}{\omega} \left[
\frac{1}{4} (\omega R)^2 (\omega T) - \frac{1}{32}  (\omega T)^3 +
\frac{9}{2048}  (\omega T)^5
 \right], \nonumber \\
\Delta \kappa &=& \frac{1}{2} (\omega R)^2 \left[\left(\frac{\Delta
x}{R}\right)^2 - \frac{1}{2}(\omega T)^2 \right] + \frac{1}{32}
(\omega T)^4 - \frac{11}{6144} (\omega T)^6. \label{Dx_Dk}
\end{eqnarray}
The first term in Eq.~(\ref{Ir}) is canceled by Eq.~(\ref{Is}) and
the incomplete overlap at the recombination time has a larger effect
than the change in the harmonics' size $R$.

Finally, $\Delta s$ is given by the expression
\begin{equation}
\Delta s = - \omega^2 \left[1 + \left( \frac{R_0}{R} \right)^3 \left(  D_3(|\Delta x|/R) - 1\right) \right],\label{Ds1}
\end{equation}
where $D_3$ is given by Eq.~(\ref{D3}).  Here, the change in the
size of the harmonics $R$ has a larger effect on $\Delta s$ than the
incomplete overlap at the recombination time.  In the limit where
the change in $R$ is small and using Eq.~(\ref{R_g_final}), reduces
Eq.~(\ref{Ds1}) to
\begin{equation}
\Delta s = - \omega^2 \frac{3}{4} (\omega T)^2.
\end{equation}

The loss of contrast due to the term $A$ Eq.~(\ref{A}) occurs when
$A > 1/2$. Using Eq.~(\ref{xpoR}) and (\ref{Dx_Dk}), we can
translate this inequality into the relation
\begin{equation}
\frac{1}{16} \frac{(\omega T)^3}{\omega R} > 1. \label{Dx_Dp}
\end{equation}
Similarly, the loss of contrast due to B Eq.~(\ref{B}) is given by
Eq.~(\ref{DkR}) and Eq.~(\ref{DeltaK}), which is the sum of three
terms.   The first term is
\begin{equation} \label{Dk}
\Delta \kappa =
\frac{1}{32} (\omega T)^4 + \frac{1}{6144} (\omega
T)^6  - \frac{3}{8} (\omega R)^2 (\omega T)^2 .
\end{equation}
 The second term is the product of the distance between the
two harmonics and the quadratic contribution to the phase at the recombination time and may be
expressed as
\begin{equation}
 g \Delta x =
\frac{1}{32} (\omega T)^4 - \frac{9}{2048}(\omega T)^6 - \frac{1}{4}
(\omega R)^2 (\omega T)^2 \label{gDx} .
\end{equation}
The third term is the cubic contribution given by
\begin{equation}
\Delta s R^{3}  = \frac{3}{4 \omega} (\omega T)^2 (\omega R)^3. \label{DsR3}
\end{equation}
Adding Eq.~(\ref{Dk}), (\ref{gDx}), and (\ref{DsR3}), the inequality
$B > 1/2$ can be translated into the relation
\begin{equation}
\left| \frac{1}{192} (\omega T)^6  - \frac{1}{14}(\omega R)^2
(\omega T)^2 \right| \frac{\omega R}{\omega} > 2.
\label{phase_dist1}
\end{equation}
The region were the loss of contrast is caused by C can be found using Eqs.~(\ref{DsR3}) and (\ref{sR3}) and results in the relation
\begin{equation} \label{spR3}
\frac{1}{\omega} (\omega T)^2 (\omega R)^3 > 80.
\end{equation}

\begin{figure}
\includegraphics[width=8.6cm]{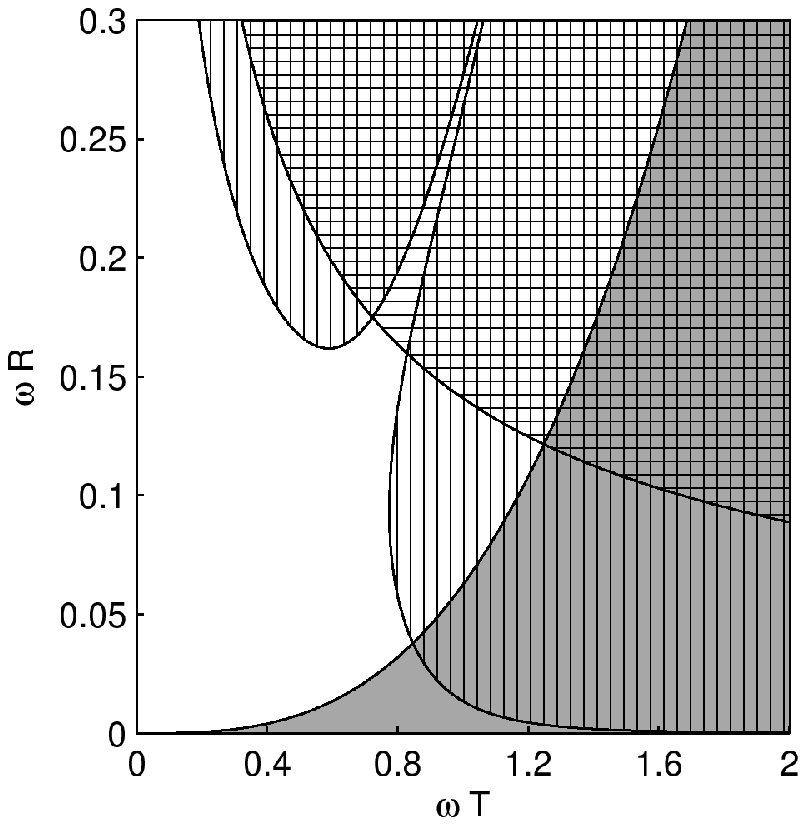}%
\caption{\label{Nominal_Map1} The regions of large and small fringe
contrast for the double reflection interferometer. The white region
corresponds to large contrast.  In the grey region $A
> 1/2$.  The region filled with vertical stripes is where $B > 1/2$.
The region filled with horizontal stripes is where $C > 1/2$. }
\end{figure}

Figure~\ref{Nominal_Map1} is a two dimensional plot showing the
regions of operation of a double reflection interferometer. The
dimensionless trap frequency is $\omega = 3.5 \times 10^{-5}$ as in
Fig. \ref{Nominal_Map_Single1}. The white region corresponds to
large contrast. In the grey region (lower right corner) $A
> 1/2$ and the contrast is small due to incomplete overlap at the
recombination time.  The region filled with vertical stripes
corresponds to $B
> 1/2$ and the contrast is small due to the phase difference across
the recombined BEC. The region filled with horizontal stripes
corresponds to $ C > 1/2$ and the cubic phase causes the loss of
contrast.

For $\omega R < 0.04$, the contrast is lost because the two
harmonics $\psi_\pm$ do not overlap at the recombination time. In
the region $0.04 < \omega R $  the contrast is lost because the trap
causes a phase difference across the recombined harmonic. Along the
curve $\omega R = \sqrt{7/96} (\omega T)^2$ the phase difference
across the recombined harmonic vanishes and $B = 0$.  However, when
$\omega T > 0.8$, the cubic phase causes the loss of contrast and $C
> 1/2$.

When the contrast is lost because of the incomplete overlap of the
clouds at the  nominal recombination time (term $A$, grey region in
Fig \ref{Nominal_Map1}), the contrast can be increased by
recombining when the two clouds overlap. Using Eqs.
(\ref{final_set_of_parab_eqns}) and (\ref{Dx_Dp}), it can be shown
that the two harmonics completely overlap at the time $\tau = T +
\Delta T$, where
\begin{equation}
\frac{\Delta T}{R} = \frac{1}{16} \frac{(\omega T)^3}{\omega R}.
\end{equation}
At this shifted time, the contrast still may be lost due to the term
$B$ if
\[
\left| \frac{1}{32} (\omega T)^4  - \frac{1}{28}(\omega R)^2 (\omega
T)^2 \right| \frac{\omega R}{ \omega} > 1
\]
and due to $C$ if Eq.~(\ref{spR3}) is fulfilled.  This method for increasing the
contrast is only useful when $\omega R < (\omega^3/2)^{1/7}$.

\begin{figure}
\includegraphics[width=8.6cm]{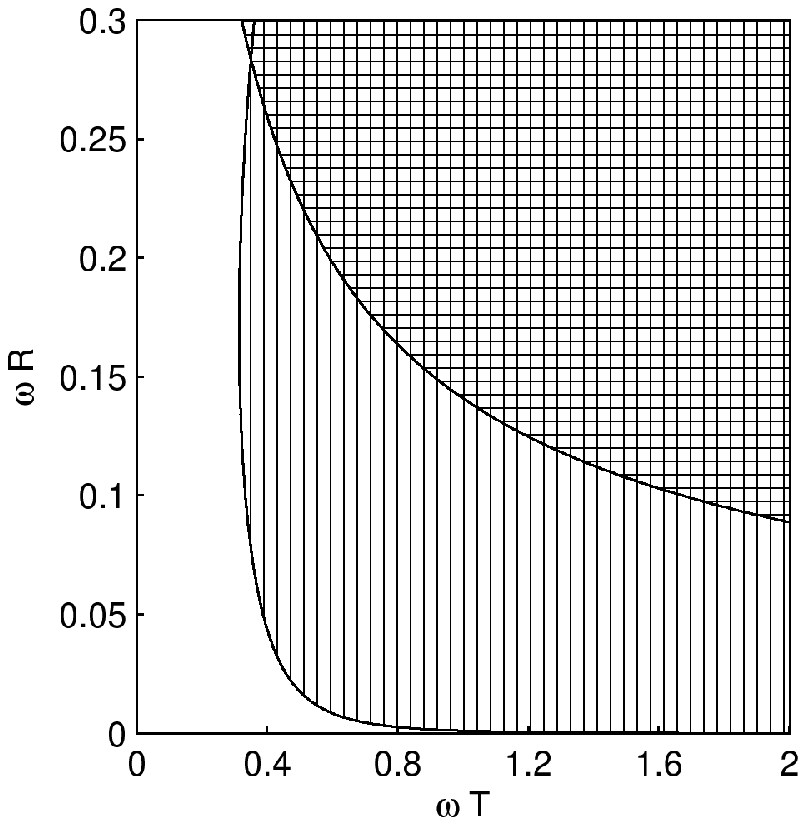}%
\caption{\label{Optimal_Map2}The regions of large and small fringe
contrast for the double reflection interferometer. The recombination
takes place at the time, when the harmonics fully overlap and $A=0$.
The white region corresponds to large contrast. The region filled
with vertical stripes is where $B
> 1/2$ and the region filled with horizontal stripes is where $C >
1/2$. }
\end{figure}

Figure \ref{Optimal_Map2} is a two-dimensional plot showing the
regions of operation when the recombination pulse is applied at the
time when the two clouds overlap.  The dimensionless trap frequency
is $3.5 \times 10^{-5}$.  The white region corresponds to large
values of the fringe contrast. The region filled with vertical
stripes corresponds to $B > 1/2$ and the region filled with
horizontal stripes to $C > 1/2$.  Recombination at a shifted time
improves the fringe contrast for $\omega R < 0.01$.

\begin{figure}
\includegraphics[width=8.6cm]{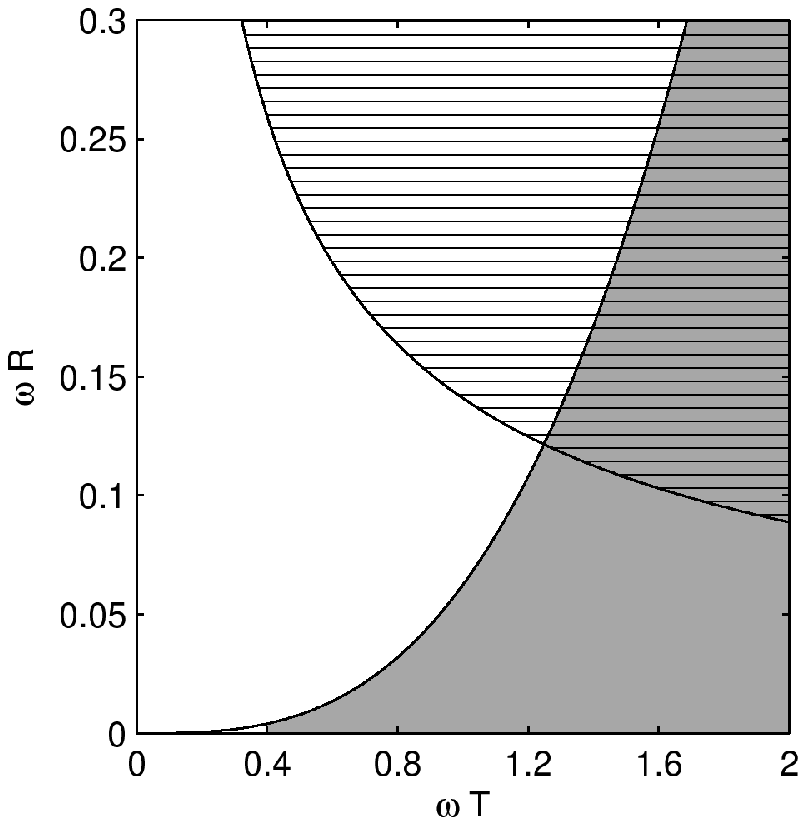}%
\caption{\label{Optimal_Map} The regions of large and small fringe
contrast for the double reflection interferometer, when the
recombination takes place at the optimal time, when $B=0$. The white
region corresponds to large values of the contrast.  In the grey
region $A > 1/2$  and in the region filled with horizontal stripes
$C > 1/2$. }
\end{figure}

When the contrast is lost because of the term $B$ (vertical stripes
in Fig.~\ref{Nominal_Map1}), the contrast also can be increased by
shifting the recombination time.  This is because the quantity that
determines the contrast in Eq.~(\ref{DkR}) is the sum of the three
terms given by Eqs. (\ref{Dk}), (\ref{gDx}), and (\ref{DsR3}). A
small change in the recombination time results in a change in
Eq.~(\ref{gDx}), but does not change either Eq.~(\ref{Dk}) or
Eq.~(\ref{DsR3}).
Using Eqs. (\ref{final_set_of_parab_eqns}) one can show that recombining at the time $\tau = T + \Delta T$, where
\begin{equation}
 \frac{\Delta T}{R} = \frac{1}{192} \frac{(\omega T)^5}{\omega R} - \frac{1}{14} (\omega T)(\omega R), \label{shifted_time}
\end{equation}
results in $B = 0$ at the shifted recombination time.
With the help of Eqs. (\ref{final_set_of_parab_eqns}),
(\ref{phase_dist1}) and (\ref{shifted_time}), the inequality $A >
1/2$ at the shifted recombination time translates into the relation
\begin{equation}
  \frac{\Delta \kappa}{g R} = \frac{1}{16} \frac{(\omega
T)^3}{\omega R_0} > 1, \label{DxoR_10}
\end{equation}
The region where $C > 1/2$ is still given by Eq.~(\ref{spR3}).
Figure \ref{Optimal_Map} is a two-dimensional plot showing the
different regions of operation when the recombination takes place at
the shifted time given by Eq.~(\ref{shifted_time}).  The
dimensionless trap frequency is taken to be $\omega = 3.5 \times
10^{-5}$.  The white region corresponds to large values of the
fringe contrast. In the grey  region the loss of contrast is caused
by the term $A$ and in the region filled with horizontal stripes by
the term $C$.
When $\omega R < 0.13$, the incomplete overlap causes the loss of
fringe contrast and when $\omega R > 0.13$ the cubic phase causes
the loss of contrast.

The lower limit on the values of $\omega R$ for which the fringe
contrast is large with the help of Eq.~(\ref{DxoR_10}) can be
written as
\begin{equation}
 \omega R \approx
\frac{1}{16} (\omega T)^3. \label{LowerLimit}
\end{equation}
and the upper limit on $\omega R$ using Eq.~(\ref{spR3}) can be
expressed as
\begin{equation}
 \omega R =  \frac{(30 \omega)^{1/3}}{(\omega T)^{2/3}}
\label{UpperLimit}.
\end{equation}
The longest cycle time before the loss of the fringe contrast is
given by the relation
\begin{equation}
\omega T_{max} \approx  3 \omega^{1/11},
\end{equation}
and occurs when the size of the BEC is
\begin{equation}
 \omega R_{opt} \approx\frac{27}{16} \omega^{3/11}.
\end{equation}

\section{Comparison with experiment}\label{experiment_comp}

There have been several experimental realizations of waveguide
interferometers that use optical pulses to control the dynamics of the BEC
\cite{wang05, horikoshi06, garcia06, burke07, deissler07}.

The JILA group \cite{wang05} built a single reflection
interferometer and used a trap with frequencies $(\omega_x,
\omega_y, \omega_z) = 2 \pi \times (100, 100, 5)$~Hz and a
$^{87}\mbox{Rb}$ BEC with about $10^4$ atoms.  These parameters
correspond to the nonlinearity parameter $p = 5.7$, the
dimensionless BEC radius $R = 675$, the dimensionless trap frequency
of $\omega = 1.7 \times 10^{-4}$ and $\omega R = 0.11$.
 In Ref. \cite{wang05} the contrast at about
10~ms was found to be $V = 0.2$. This agrees with our model if,
instead of using Eqs.~(\ref{SingleReflect}) and
Eq.~(\ref{contrast_final}), we use Eq.~(\ref{SingleReflect}) and
numerically integrate Eq.~(\ref{contrast}).

In the experiment at the University of Virginia \cite{burke07} the
trap has frequencies $(\omega_x, \omega_y, \omega_z) = 2 \pi \times
(3.3, 6, 1.2)$~Hz, and the BEC had about $3 \times 10^4$
$^{87}\mbox{Rb}$ atoms. These parameters correspond to the
dimensionless nonlinearity parameter  $p \approx 0.75$, the
dimensionless BEC radius of $R = 940$, the dimensionless trap
frequency of $\omega = 3.65 \times 10^{-5}$ and $\omega R = 0.035$.

The University of Virginia group experimentally investigated the
operation of both a single and double reflection interferometer.
When analyzing the single reflection interferometer, they found 50\%
contrast for the cycle time of about 12 ms. This time corresponds to
$\omega T = 0.9$ (cf. Eq.(\ref{omegaT})).  The loss of contrast at
this time agrees with Fig.~\ref{Nominal_Map_Single1} and the results
of Sec. \ref{Single_Reflect}.

The double reflection interferometer had 50\% contrast for a cycle
time of 80 ms, translating to $\omega T = 0.6$ in our variables. The
loss of contrast at this time agrees with Fig.~\ref{Nominal_Map1}
and the results of Sec. \ref{Double_Reflect}. The experiment used a
BEC radius $R$ such that the contrast was lost due to both the
incomplete overlap (term $A$) and the phase across the cloud (term
$B$) at about the same cycle time $\omega T$. This may explains why
the University of Virginia experiment had no phase distortion across
the recombined BEC. An increase of decrease in the radius $R$ would
have caused a decrease in the coherence time of the interferometer.
However, our model predicts that by both increasing the radius and
shifting the recombination time it would be possible to increase the
coherence time.

\section{Conclusions} \label{Conclusions}

In this paper, we analyzed the operation of both single and double
reflection interferometers. We introduced a simple analytic model to
determine the regions in parameter space where the fringe contrast
is large and small. For the case of a double reflection
interferometer, we showed that the coherence time can  be increased
by changing the recombination time. Finally, we compared our results
to recent experimental realizations of these interferometers.

Our analysis focused on the case where the BEC was in the ground
state of the trap at the beginning of the interferometric cycle.
Analysis of the single reflection interferometer when this
restriction is relaxed can be found in \cite{stickney07}.   In the
case of a double reflection interferometer, the two largest terms
(of order $(\omega T)^4$) in Eqs. (\ref{Dk}) and (\ref{gDx}) are
only equal when the BEC is initially in the ground state.  As a
result, the region of high contrast discussed in Sec.
\ref{Double_Reflect} becomes smaller when the BEC is not initially
in the ground state.

The analysis of this paper did not include effects beyond the mean
field approximation such as phase diffusion and finite temperature
phase fluctuations.

The phase fluctuations of a BEC in a trap have been extensively
studied in Ref.~\cite{lewenstein96, javanainen97, leggett98,
javanainen98}. It has been shown that the phase diffusion time  has
the functional form $T_D \sim (\bar a_{HO} / a_s)^{2/5} N^{1/10}  /
{\bar \omega}$ \cite{javanainen97}, where $\bar \omega = (\omega_x
\omega_y \omega_z)^{1/3}$ is the geometric average of the trap
frequencies and $\bar a_{HO} = \sqrt{\hbar / M \bar \omega}$ is the
average oscillator length.  For the case of the University of
Virginia experiment \cite{burke07}, the phase diffusion time is $T_D
\sim 3~\mbox{sec}$, which is much longer than the upper limit
calculated in Sec. \ref{experiment_comp}.

When the aspect ratio of the trap $\omega_{\parallel}/\omega_\perp$
becomes sufficiently large, the BEC becomes one-dimensional and
phase fluctuations can cause decoherence \cite{petrov01, dettmer01,
burkov07, hofferberth07}.  The  phase fluctuations along the BEC
depend on the aspect ratio and temperature of the BEC and become
important when the temperature of the BEC is larger than $T = 15
(\hbar \omega_\parallel)^2 N / 32 \mu$ \cite{petrov01}, where $\mu$
is the chemical potential of the BEC.  For the recent experiments
\cite{wang05, garcia06}, the phase fluctuations across the BEC are
sufficiently small that they can be neglected.

\section{Acknowledgements}
This work was supported by the Defense Advanced Research Projects
Agency (Grant No. W911NF-04-1-0043) and the National Science
Foundation (Grant No. PHY-0551010).

\end{document}